\documentclass[8.5pt,twoside,twocolumn]{article}

\usepackage{mhchem}
\usepackage{times,mathptmx}
\usepackage{sectsty}
\usepackage{balance} 

\usepackage{graphicx} %eps figures can be used instead
\usepackage{lastpage}
\usepackage[format=plain,justification=raggedright,singlelinecheck=false,font=small,labelfont=bf,labelsep=space]{caption} 
\usepackage{fancyhdr}
\begin{document}

\thispagestyle{plain}
\fancypagestyle{plain}{
%%%\fancyhead[L]{\includegraphics[height=8pt]{headers/LH}}
%%%\fancyhead[C]{\hspace{-1cm}\includegraphics[height=20pt]{headers/CH}}
%%%\fancyhead[R]{\includegraphics[height=10pt]{headers/RH}\vspace{-0.2cm}}
\renewcommand{\headrulewidth}{1pt}}
\renewcommand{\thefootnote}{\fnsymbol{footnote}}
\renewcommand\footnoterule{\vspace*{1pt}% 
\hrule width 3.4in height 0.4pt \vspace*{5pt}} 
\setcounter{secnumdepth}{5}

\makeatletter 
\def\subsubsection{\@startsection{subsubsection}{3}{10pt}{-1.25ex plus -1ex minus -.1ex}{0ex plus 0ex}{\normalsize\bf}} 
\def\paragraph{\@startsection{paragraph}{4}{10pt}{-1.25ex plus -1ex minus -.1ex}{0ex plus 0ex}{\normalsize\textit}} 
\renewcommand\@biblabel[1]{#1}            
\renewcommand\@makefntext[1]% 
{\noindent\makebox[0pt][r]{\@thefnmark\,}#1}
\makeatother 
\renewcommand{\figurename}{\small{Fig.}~}
\sectionfont{\large}
\subsectionfont{\normalsize} 

\fancyfoot{}
%%%\fancyfoot[LO,RE]{\vspace{-7pt}\includegraphics[height=9pt]{headers/LF}}
%%%\fancyfoot[CO]{\vspace{-7.2pt}\hspace{12.2cm}\includegraphics{headers/RF}}
%%%\fancyfoot[CE]{\vspace{-7.5pt}\hspace{-13.5cm}\includegraphics{headers/RF}}
%%%\fancyfoot[RO]{\footnotesize{\sffamily{1--\pageref{LastPage} ~\textbar  \hspace{2pt}\thepage}}}
%%%\fancyfoot[LE]{\footnotesize{\sffamily{\thepage~\textbar\hspace{3.45cm} 1--\pageref{LastPage}}}}
\fancyhead{}
\renewcommand{\headrulewidth}{1pt} 
\renewcommand{\footrulewidth}{1pt}
\setlength{\arrayrulewidth}{1pt}
\setlength{\columnsep}{6.5mm}

\twocolumn[
  \begin{@twocolumnfalse}
\noindent\Large{\textbf{Traces of Radioactive $^{131}$I in Rain Water and Milk Samples in Romania}}
\vspace{0.6cm}

\noindent\large{\textbf{Romul Margineanu,$^{\ast}$\textit{$^{a}$} Bogdan Mitrica,$^{\ast}$\textit{$^{a}$} Ana Apostu,$^{\ast}$\textit{$^{a}$}
 and
Claudia Gomoiu$^{\ast}$\textit{$^{a}$}}}\vspace{0.5cm}
%Please note that \ast indicates the corresponding author(s) but no footnote text is required. 

%%%\noindent\textit{\small{\textbf{Received Xth XXXXXXXXXX 20XX, Accepted Xth %%%XXXXXXXXX 20XX\newline
%%%First published on the web Xth XXXXXXXXXX 200X}}}
%%%
%%%\noindent \textbf{\small{DOI: 10.1039/b000000x}}
\vspace{0.6cm}
%Please do not change this text.

\noindent \normalsize{{\bf Abstract} Measurements of $^{131}$I (T$_{1/2}$ = 8.04 days) have been performed in IFIN-HH's underground laboratory situated in Unirea salt mine from Slanic-Prahova, Romania. The rain water samples were collected in March 27$^{th}$ from Brasov and in March 27$^{th}$, 29$^{th}$ and April 2$^{nd}$ from Slanic-Prahova. Also sheep milk was collected in Slanic area and subsequently measured. The samples were measured in the IFIN-HH's underground laboratory, in ultra-low radiation background, using a high resolution gamma-ray spectrometer equipped with a GeHP detector having a FWHM = 1.80 keV at 1332.48 keV at the second $^{60}$Co gamma-ray and a relative efficiency of 22.8 \%. The results show a specific activity of $^{131}$I from 0.15 to 0.75 Bq/dm$^{3}$ for rains. In the sheep milk from Slanic area the specific activity of $^{131}$I was about 5.2 Bq/dm$^{3}$.}
\vspace{0.5cm}
 \end{@twocolumnfalse}
  ]

\section{Introduction}
%Footnotes
%%%\footnotetext{\dag~Electronic Supplementary Information (ESI) available: [details of any supplementary information available should be included here]. See DOI: 10.1039/b000000x/}

%Please use \dag to cite the ESI in the main text of the article.
%If you article does not have ESI please remove the the \dag symbol from the title and the above footnotetext.

\footnotetext{\textit{$^{a}$~Horia Hulubei National Institute of Physics and Nuclear Engineering - IFIN HH, Reactorului 30, P.O.BOX MG-6, 077125, Magurele, Ifov, Romania, romulus@ifin.nipne.ro, mitrica@ifin.nipne.ro, anapostu@ifin.nipne.ro, cgomoiu@ifin.nipne.ro}
}

%additional addresses can be cited as above using the lower-case letters, c, d, e... If all authors are from the same address, no letter is required

%%%\footnotetext{\ddag~Additional footnotes to the title and authors can be included \emph{e.g.}\ `Present address:' or `These authors contributed equally to this work' as above using the symbols: \ddag, \textsection, and \P. Please place the appropriate symbol next to the author's name and include a \texttt{\textbackslash footnotetext} entry in the the correct place in the list.}

%%The main text of the article\cite{Mena2000} should appear here.

The Fukushima accident started on March 11$^{th}$, 2011 causes the release of significant amounts of $^{131}$I, $^{171}$Cs and other radioactive isotopes in the environment. The atmospheric currents spread the radioactive contamination all over northern hemisphere. According with meteorological information the radioactive cloud has reached the Romanian territory beginning with March 25-26, \cite{france, protv}. The meteorological conditions over Romania were characterised by small rains. Six sample, five of rain water and one of sheep milk, were taken for analyse.

\begin{figure}[h]
\centering
  \includegraphics[height=3cm]{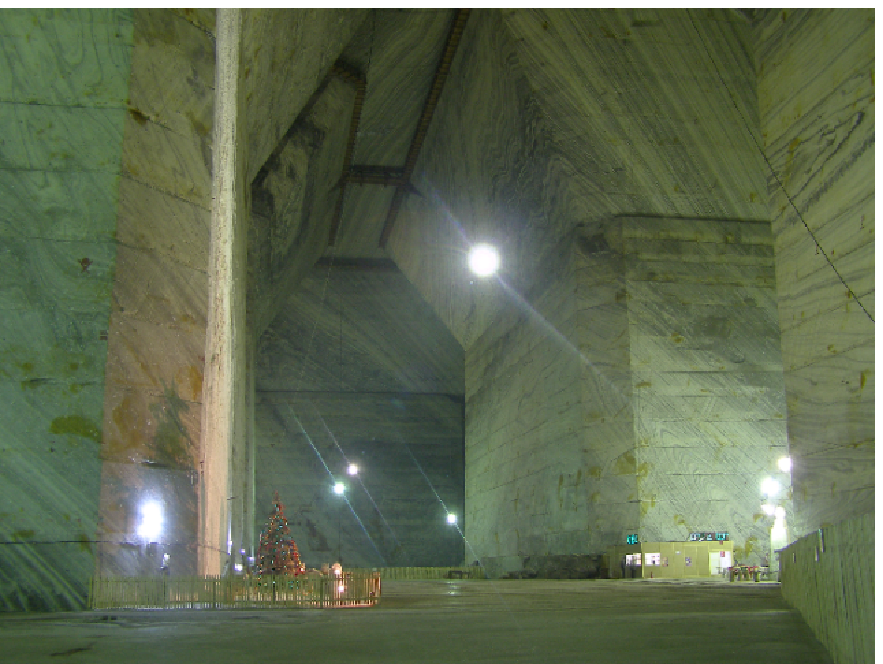}
  \includegraphics[height=3cm]{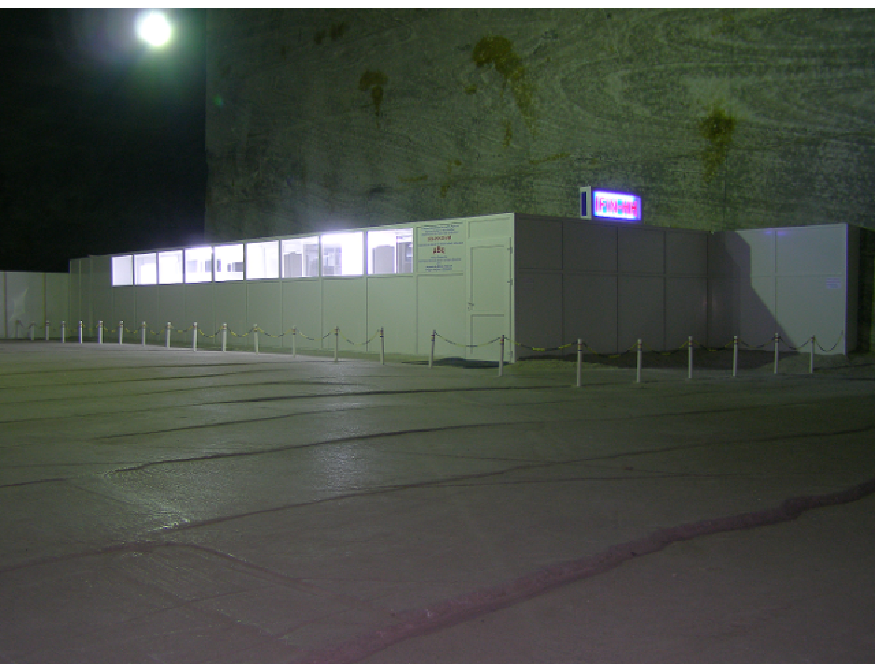}
  \caption{Photo from Unirea salt mine – left, photo of the underground laboratory - right}
  \label{mine}
\end{figure}

The samples were measured in the institute's underground laboratory from Slanic-Prahova, see Fig. \ref{mine}. The laboratory was constructed and putted in operation in 2006, \cite{margineanu1}. The characteristics of the galleries of the Unirea salt mine are:

- depth: 208 m bellow ground level

- temperature: 12.0 -13.0 $^{\circ}$C

- humidity: 65-70 \% 

- excavated volume: 2.9 million m$^{3}$ 

- floor area: 70000 m$^{2}$

- average high: 52-57 m

- distance between walls: 32-36 m

- existing infrastructure: electricity, elevator, phone, Internet, GSM networks.

- equivalent depth – from cosmic ray muon measurements: 610 mwe (meter water equivalent), \cite{mitrica}.

\section{Measurements and results}
\begin{figure}[h]
\centering
  \includegraphics[height=5cm]{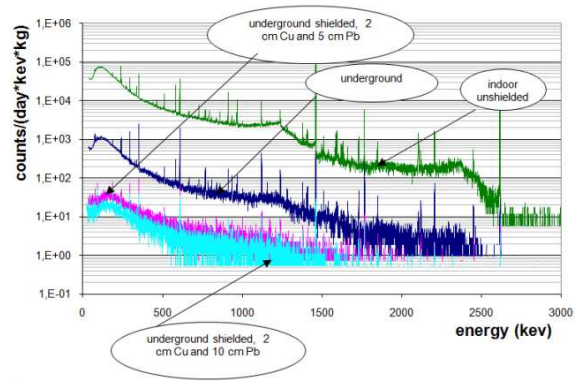}
  \caption{Four experimental gamma spectra of the background measured indoor at surface and in underground – unshielded, shielded with 5 cm Lead and 2 cm Copper and shielded with 10 cm Lead and 2 cm Copper \cite{margineanu2}}
 \label{background}
\end{figure}

The measurements were performed with a CANBERRA ultra-low GeHP system, equipped with a detector having a relative efficiency of 22.8 \%, assisted by an INSPECTOR 2000 multichannel analyser and GENIE 2000 software code. The detector is housed in a 10 cm Lead and 2 cm Copper shield, which assures a reduction of the background of 1600 times with respect to the spectrum collected outdoor at surface, see Fig. \ref{background}.

\begin{figure}[h]
\centering
  \includegraphics[height=5cm]{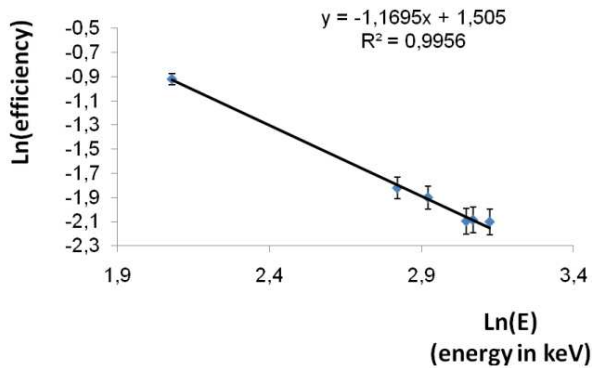}
  \caption{Efficiency versus energy for the spectrometric system equipped with an ultra-low GeHP detector with 22.8 \% relative efficiency}
  \label{efficiency}
\end{figure}

The efficiency of the measurement system was determined with IAEA-444 reference material which is a soil from China containing a cocktail of $^{109}$Cd, $^{60}$Co, $^{137}$Cs, $^{54}$Mn and $^{65}$Zn radionuclides. The energy dependence of efficiency is represented in log-log graph in Fig. \ref{efficiency}. 

\begin{figure}[h]
\centering
  \includegraphics[height=10cm]{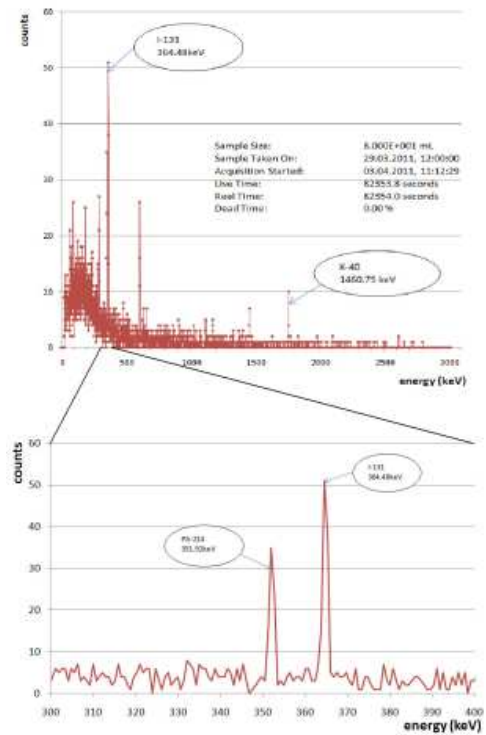}
  \caption{Gamma ray spectrum of $^{131}$I in the rain water from Slanic collected in the morning of 29 March 2011 – up, a detail of the same spectrum -down}
  \label{rain}
\end{figure}

Samples of rain water were collected beginning with March 27 and were measured in cylindrical plastic box of 75 mm diameter and 35 mm high. The volume of measured samples was 80 cm3. The $^{131}$I line at 364.48 keV has been seen in all collected spectra. This aspect is illustrated for rain water sample collected in the morning of March 29$^{th}$ in Slanic in Fig. \ref{rain}. The specific activities of $^{131}$I in the rain water samples are presented in Tab. \ref{iod}.

%%%%%%%%%%%% tabular %%%%%%%%%%%%
\begin{table}[h]
\small
  \caption{\ Specific activity of $^{131}$I  in rain water samples}
  \label{iod}
  \begin{tabular*}{0.5\textwidth}{@{\extracolsep{\fill}}llll}
    \hline
    Sample & Location & Sampling date  & Bq/dm$^{3}$\\
    \hline
     1 & Brasov & March 27, 2011 & 0.41 $\pm$ 0.04\\
     2 & Slanic  & March 27, 2011 - morning & 0.52 $\pm$ 0.05\\
     3 & Slanic  & March 27, 2011 - evening & 0.15 $\pm$ 0.02\\
     4 & Slanic  & March 29, 2011 & 0.75 $\pm$ 0.06\\
     5 & Slanic  & April 2, 2011 & 0.69 $\pm$ 0.06\\
    \hline
  \end{tabular*}
\end{table}

In April 5th a sheep milk sample has been collected and subsequently measured in the same way as rain water samples. The specific activity of $^{131}$I measured in sheep milk is 5.2 $\pm$ 0.5 Bq/dm$^{3}$.

\section{Comments} 
From the results, one can observe the presence of $^{131}$I in very small amounts in the precipitation and milk recorded beginning with 27 March 2011 in Brasov and Slanic Prahova, Romania. The specific activity in rain water of $^{131}$I varies from 0,15 Bq/m$^{2}$ to 0,75 Bq/m$^{2}$.

For the moment, no other data have been available for analyses, but even so we can suppose that the $^{131}$I originates from Fukushima nuclear accident. For this reason, the environmental radioactivity, especially in rain water and milk, is monitored continuously in order to assess the level of radioactive deposition. 

The measured activities are far bellow any intervention limits. For instance in Japan the limit for drinking water was set at 300 Bq/dm$^{3}$ for adults and children and 100 Bq/dm$^{3}$ for infants \cite{riken}. The values measured by us are 2 to 3 order or magnitude lower than these limits. In sheep milk the $^{131}$I concentration is more than an order of magnitude lower than the limits.

The presence of $^{131}$I over Romania demonstrates that the consequences of a nuclear accident could be put in to evidence even at more than 10,000 km away which also demonstrates that at this moment the radioactive plume originating in Fukushima is spread practically all over the Northern Hemisphere.

% an example of a two-column figure
%\begin{figure*}
  %\centering
  %\includegraphics[height=3cm]{example.jpg}
  %\caption{An example figure caption, an image from the \textit{Physical Chemistry Chemical Physics} cover gallery.}
  %\label{fgr:example}
%\end{figure*}

% an example of a two-column table
%\begin{table*}
%\small
  %\caption{\ An example of a caption to accompany a table, table captions do not end in a full point}
  %\label{tbl:example}
  %\begin{tabular*}{\textwidth}{@{\extracolsep{\fill}}lllllll}
    %\hline
    %Header one & Header two & Header three & Header four & Header five & Header six  & Header seven\\
    %\hline
    %1 & 2 & 3 & 4 & 5 & 6  & 7\\
    %8 & 9 & 10 & 11 & 12 & 13 & 14 \\
    %15 & 16 & 17 & 18 & 19 & 20 & 21\\
    %\hline
  %\end{tabular*}
%\end{table*}

%The \balance command can be used to balance the columns on the final page if desired. It should be placed anywhere within the first column of the last page.

%\balance

%If notes are included in your references you can change the title from 'References' to 'Notes and references' using the following command:
%\renewcommand\refname{Notes and references}

%%%\footnotesize{
%%%\bibliography{mitrica} %your .bib file
%%%\bibliographystyle{rsc} %the RSC's .bst file
%%%}

\footnotesize{

}

\end{document}